\documentclass{iopart}
\usepackage{amsmath,amssymb}
\usepackage{tikz}
\usetikzlibrary{shapes}

\DeclareMathAccent{\wtilde}{\mathord}{largesymbols}{"65}
\DeclareMathAccent{\what}{\mathord}{largesymbols}{"62}

\def\m@th{\mathsurround=0pt}
\mathchardef\bracell="0365 
\def\upbrall{$\m@th\bracell$}
\def\undertilde#1{\mathop{\vtop{\ialign{##\crcr
    $\hfil\displaystyle{#1}\hfil$\crcr
     \noalign
     {\kern1.5pt\nointerlineskip}
     \upbrall\crcr\noalign{\kern1pt
   }}}}\limits}
\def\theequation{\arabic{section}.\arabic{equation}}

\newcommand{\bde}{\boldsymbol{e}}

\newcommand{\bn}{\boldsymbol{n}}

\newcommand{\bLm}{\boldsymbol{\Lambda}}

\newcommand{\al}{\alpha}
\newcommand{\bt}{\beta}
\newcommand{\gm}{\gamma}

\newcommand{\dl}{\delta}
\newcommand{\Dl}{\Delta}

\newcommand{\cL}{\mathcal{L}}

\newcommand{\Li}{{\rm Li}}

\newcommand{\bblu}{\begin{color}{blue}}
\newcommand{\bred}{\begin{color}{red}}
\newcommand{\ecl}{\end{color}}

\newcommand{\nn}{\nonumber}

\newcommand{\be}{\begin{equation}}
\newcommand{\ee}{\end{equation}}
\newcommand{\bea}{\begin{eqnarray}}
\newcommand{\eea}{\end{eqnarray}}
\newcommand{\bse}{\begin{subequations}}
\newcommand{\ese}{\end{subequations}}

\begin{document}
\title{Lagrangian multiform structure for the lattice KP system} 
\author{S.B. Lobb$^1$, F.W. Nijhoff$^1$ and G.R.W. Quispel$^2$}
\address{
$^1$ Department of Applied Mathematics, University of Leeds, Leeds LS2 9JT, UK\\
$^2$ Department of Mathematics and Statistics, La Trobe University, Victoria 3086, Australia}

\begin{abstract}
We present a Lagrangian for the bilinear discrete KP (or Hirota-Miwa) equation. Furthermore, we show that this Lagrangian can be extended to a Lagrangian 3-form when embedded in a higher dimensional lattice, obeying a closure relation. Thus we establish the multiform structure as proposed in \cite{closure} in a higher dimensional case.
\end{abstract}

\section{Introduction}

In \cite{closure} the idea was put forward that lattice systems which are integrable in the sense of multidimensional consistency \cite{CAC,B+S} should have a Lagrangian structure which reflects this property. That is, rather than the Lagrangian being a scalar object (or equivalently a volume form), it should be a discrete multiform from which, through the Euler-Lagrange equations, copies of the relevant equation in all possible lattice directions can be derived. These copies of the same equation, albeit with different parameters associated with different lattice directions, coexist on an extended lattice in view of the multidimensional consistency, and should consequently be viewed as parts of one single ``integrable'' infinite-dimensional system. Examples from a particular class of quadrilateral lattice systems in 1+1 dimensions (those classified in \cite{ABS_quad}) were studied in \cite{closure}, namely equations of the form
\be
 Q(u,u_{i},u_{j},u_{ij};\al_{i},\al_{j})=0,
\ee
where $u=u(n_{i},n_{j})$ depends on two discrete variables $n_{i},n_{j}$, shifts of $u$ in the $n_{i}$-direction are denoted by $u_{i}$ (so that for example $u_{i}=u(n_{i}+1,n_{j})$), and the $\al_{i}$ are lattice parameters associated with the $n_{i}$-direction. Although actions for these equations were given in \cite{ABS_quad}, it was shown in \cite{closure} that all cases admit a special choice of 3-point Lagrangians, which subsequently can be interpreted as Lagrangian 2-forms. This was based on the surprising observation that such Lagrangians obey the following closure relation
\be\label{2dclos}
 \Dl_{i}\cL_{jk}+\Dl_{j}\cL_{ki}+\Dl_{k}\cL_{ij} = 0,
\ee
which implies they are closed 2-forms on a multidimensional lattice. Here the difference operator $\Dl_{i}$ acts on functions $f$ of $u=u(n_{i},n_{j},n_{k})$ by the formula $\Dl_{i}f(u)=f(u_{i})-f(u)$, and on a function $g$ of $u$ and its shifts by the formula $\Dl_{i}g(u,u_{j},u_{k})=g(u_{i},u_{ij},u_{ik})-g(u,u_{j},u_{k})$. On the basis of the relation \eqref{2dclos} a new variational principle was proposed for integrable (in the sense of multidimensional consistency) lattice equations which involves the geometry of the space of independent variables.

\paragraph{}
Whereas in the previous paper the focus was on integrable lattice equations in 1+1 dimensions, here we want to study the case of 3-dimensional integrable systems, the prime example being the lattice Kadomtsev-Petviashvili (KP) system. Discrete equations of KP type have been studied extensively since the early 1980s (cf for example \cite{DJM,Frank1984}), following on from the famous ``discrete analogue of a generalized Toda equation'' (DAGTE) introduced by Hirota in \cite{Hirota} which is a bilinear form for the lattice KP equation\footnote{Hirota introduced his difference equation in a form equivalent to
\be\label{Hirota_eqn}
\al\tau_{i}\tau_{\bar{\imath}}+\bt\tau_{j}\tau_{\bar{\jmath}}+\gm\tau_{k}\tau_{\bar{k}}=0,
\ee
where the notation is explained in the text, and where $\al,\bt,\gm$ are constants satisfying $\al+\bt+\gm=0$.}. Other related KP-type lattice equations were introduced in \cite{Frank1984}. The equation we will refer to as the bilinear discrete KP equation, in order to distinguish it from equations that actually lead to the original KP equation in a continuum limit, is taken in the following form
\be\label{AKP}
A_{jk}\tau_{i}\tau_{jk}+A_{ki}\tau_{j}\tau_{ki}+A_{ij}\tau_{k}\tau_{ij}=0.
\ee
Here $A_{ij}=-A_{ji}$ are constants, $\tau=\tau(n_{i},n_{j},n_{k})$ is the dependent variable depending on three discrete independent variables $n_{i},n_{j},n_{k}$ corresponding to lattice directions, and subscripts of $\tau$, e.g. as in $\tau_{i}$, denote shifts in the $n_{i}$-direction so that for example $\tau_{i}=\tau(n_{i}+1,n_{j},n_{k})$ and $\tau_{\bar{\jmath}}=\tau(n_{i},n_{j}-1,n_{k})$. The constants can be removed by a gauge transformation, but we find it more instructive to retain them. Miwa gave the connection between the KP hierarchy and Hirota's difference equation in \cite{Miwa}, showing how solutions to the KP hierarchy can be transformed into solutions to \eqref{AKP}, hence it is often called the \emph{Hirota-Miwa} equation.

\paragraph{}
The main results of this paper are twofold: first to give a Lagrangian for the bilinear discrete KP system associated with \eqref{AKP} (in fact, whereas the continuous KP equation admits an obvious Lagrangian structure, it has to our knowledge never been established for any KP-type equation on the 3-dimensional lattice), second to establish the Lagrangian multiform structure, in the sense of \cite{closure}, based on a higher dimensional analogue of \eqref{2dclos}, and show that the relevant Lagrangian obeys a 4-dimensional closure relation.

\section{Lagrangian structure}
\subsection{Scalar Lagrangian}
It is a common feature of Lagrangians for equations of Korteweg-de Vries (KdV) and KP type (already in the continuous case) that those equations emerge as Euler-Lagrange equations by varying the action with respect to a dependent variable which obeys a potential (i.e. integrated) version of the equation. Hence, the variational equation is typically a ``derived form'' of the equation obeyed by this canonical variable, with respect to which the action is minimized. The same holds true in the case of a Lagrangian structure for the lattice KP system, where we will use the $\tau$-function as the canonical variable. Thus, fixing three directions $i,j,k$, we introduce the following Lagrangian 
\bea\label{F123}
 L(\tau_{i},\tau_{j},\tau_{k},\tau_{ij},\tau_{jk},\tau_{ki};A_{ij},A_{jk},A_{ki})&&\nn\\
            = \ln\biggl(\frac{\tau_{k}\tau_{ij}}{\tau_{j}\tau_{ki}}\biggr)\ln\biggl(-\frac{A_{ki}\tau_{j}}{A_{jk}\tau_{i}}\biggr)
                      -\Li_{2}\biggl(-\frac{A_{ij}\tau_{k}\tau_{ij}}{A_{ki}\tau_{j}\tau_{ki}}\biggr) & =: & L_{ijk},
\eea
where $\Li_{2}$ denotes the dilogarithm function defined by
\be\label{dilog}
 \Li_2(z) = -\int^z_0{\frac{\ln(1-z)}{z}dz}.
\ee
The Lagrangian \eqref{F123} produces the following discrete Euler-Lagrange equation
\bea\label{4copies}
\fl \frac{\dl L}{\dl\tau} & = & \biggl\{
            \ln\biggl(-\frac{A_{ki}\tau_{j\bar{k}}\tau_{i}+A_{ij}\tau\tau_{ij\bar{k}}}{A_{jk}\tau_{i\bar{k}}\tau_{j}}\biggr)
           +\ln\biggl(-\frac{A_{ki}\tau_{\bar{\jmath}k}\tau_{\bar{\imath}}+A_{ij}\tau\tau_{\bar{\imath}\bar{\jmath}k}}
                            {A_{jk}\tau_{\bar{\imath}k}\tau_{j}}\biggr)\nn\\
\fl &&     -\ln\biggl(-\frac{A_{ki}\tau\tau_{i\bar{\jmath}k}+A_{ij}\tau_{\bar{\jmath}k}\tau_{i}}{A_{jk}\tau_{i\bar{\jmath}}\tau_{k}}\biggr)
           -\ln\biggl(-\frac{A_{ki}\tau\tau_{\bar{\imath}j\bar{k}}+A_{ij}\tau_{j\bar{k}}\tau_{\bar{\imath}}}
                            {A_{jk}\tau_{\bar{\imath}j}\tau_{\bar{k}}}\biggr)\biggr\}\frac{1}{\tau}\nn\\
\fl & = & 0
\eea
which is a consequence of \eqref{AKP} through the fact that it is a combination of 4 copies of the equation shifted in appropriate lattice directions.

Consequently the following functional of the lattice fields $\tau(n_{i},n_{j},n_{k})$
\be 
S[\tau]=\sum_{n_{i},n_{j},n_{k}}L(\tau_{i},\tau_{j},\tau_{k},\tau_{ij},\tau_{jk},\tau_{ki};A_{ij},A_{jk},A_{ki})
\ee
with $L$ given by \eqref{F123} can be considered to constitute an action for the lattice equation \eqref{4copies} as a derived equation of the bilinear discrete KP equation. However, we want to go further and take into account that the bilinear KP equation is part of a multidimensionally consistent system of equations, as has been recognized in recent years, cf e.g. \cite{Zabrodin,ABS_oct,PapNijQui}. In order to incorporate this multidimensionally consistent system of equations into a single Lagrangian framework we will now proceed to define the Lagrangian multiform structure for the lattice KP system.

\subsection{Lagrangian 3-form}
The first step is to introduce a Lagrangian 3-form $\cL_{ijk}$ where $i,j,k$ denote any three distinct directions in a multidimensional lattice $\bLm$, whose vertices are labelled by integer vectors $\bn=(n_{i})_{i\in I}$ where $I$ is an arbitrary set of labels, $i,j,k$ taking values in $I$.
The lattice 3-form $\cL_{ijk}$ is based on the form of the Lagrangian \eqref{F123}, but we require it to be skewsymmetric (i.e. antisymmetric with respect to the swapping of any two indices) and we associate with it an elementary oriented cube $\sigma_{ijk}$ spanned by unit vectors $\bde_{i}$ which are associated with the corresponding lattice direction labelled by $i$ in the multidimensional lattice $\bLm$. This leads us to define the following Lagrangian 3-form

\bea
 \cL_{ijk} & = & \frac{1}{2}\bigl(L_{ijk}+L_{jki}+L_{kij}-L_{ikj}-L_{jik}-L_{kji}\bigr)\nn
\eea
which when written out explicitly and simplified is

\bea\label{L123}
\fl \cL_{ijk} & = & \;\;\;\;\ln\biggl(\frac{\tau_{k}\tau_{ij}}{\tau_{j}\tau_{ki}}\biggr)\ln\biggl(-\frac{A_{ki}\tau_{j}}{A_{jk}\tau_{i}}\biggr)
                    -\Li_{2}\biggl(-\frac{A_{ij}\tau_{k}\tau_{ij}}{A_{ki}\tau_{j}\tau_{ki}}\biggr)\nn\\
\fl              && +\ln\biggl(\frac{\tau_{i}\tau_{jk}}{\tau_{k}\tau_{ij}}\biggr)\ln\biggl(-\frac{A_{ij}\tau_{k}}{A_{ki}\tau_{j}}\biggr)
                    -\Li_{2}\biggl(-\frac{A_{jk}\tau_{i}\tau_{jk}}{A_{ij}\tau_{k}\tau_{ij}}\biggr)\nn\\
\fl              && +\ln\biggl(\frac{\tau_{j}\tau_{ki}}{\tau_{i}\tau_{jk}}\biggr)\ln\biggl(-\frac{A_{jk}\tau_{i}}{A_{ij}\tau_{k}}\biggr)
                    -\Li_{2}\biggl(-\frac{A_{ki}\tau_{j}\tau_{ki}}{A_{jk}\tau_{i}\tau_{jk}}\biggr)\nn\\
\fl              && -\frac{1}{2}\bigl(\bigl(\ln\bigl(\tau_{ij}\bigr)\bigr)^2+\bigl(\ln\bigl(\tau_{jk}\bigr)\bigr)^2
                                     +\bigl(\ln\bigl(\tau_{ki}\bigr)\bigr)^2-\bigl(\ln\bigl(\tau_{i}\bigr)\bigr)^2
                                     -\bigl(\ln\bigl(\tau_{j}\bigr)\bigr)^2-\bigl(\ln\bigl(\tau_{k}\bigr)\bigr)^2\nn\\
\fl              && \;\;\;\;\;\;\;\; -\ln\bigl(\tau_{ij}\bigr)\ln\bigl(\tau_{jk}\bigr)-\ln\bigl(\tau_{jk}\bigr)\ln\bigl(\tau_{ki}\bigr)
                                     -\ln\bigl(\tau_{ki}\bigr)\ln\bigl(\tau_{ij}\bigr)+\ln\bigl(\tau_{i}\bigr)\ln\bigl(\tau_{j}\bigr)\nn\\
\fl              && \;\;\;\;\;\;\;\; +\ln\bigl(\tau_{j}\bigr)\ln\bigl(\tau_{k}\bigr)+\ln\bigl(\tau_{k}\bigr)\ln\bigl(\tau_{i}\bigr)
                                     +(\ln(A_{ij}))^2+(\ln(A_{jk}))^2+(\ln(A_{ki}))^2\nn\\
\fl              && \;\;\;\;\;\;\;\; -\ln(A_{ij})\ln(A_{jk})-\ln(A_{jk})\ln(A_{ki})-\ln(A_{ki})\ln(A_{ij})+\frac{\pi^2}{2}\bigr),
\eea
where the constant terms arise from dilogarithm identities which will be elucidated in the proof below.

This Lagrangian is antisymmetric by construction. Considered as a usual scalar Lagrangian defined in the 3-dimensional sublattice of the directions $i,j,k$ the Euler-Lagrange equations of the corresponding action would yield an equation combining 12 shifted copies of the original bilinear equation \eqref{AKP}, namely

\bea\label{12copies}
\fl \frac{\dl\cL_{ijk}}{\dl\tau} & = & \biggl\{
            \ln\biggl(-\frac{A_{ki}\tau_{j\bar{k}}\tau_{i}+A_{ij}\tau\tau_{ij\bar{k}}}{A_{jk}\tau_{i\bar{k}}\tau_{j}}\biggr)
           +\ln\biggl(-\frac{A_{ki}\tau_{\bar{\jmath}k}\tau_{\bar{\imath}}+A_{ij}\tau\tau_{\bar{\imath}\bar{\jmath}k}}
                            {A_{jk}\tau_{\bar{\imath}k}\tau_{j}}\biggr)\nn\\
\fl  &&    -\ln\biggl(-\frac{A_{ki}\tau\tau_{i\bar{\jmath}k}+A_{ij}\tau_{\bar{\jmath}k}\tau_{i}}{A_{jk}\tau_{i\bar{\jmath}}\tau_{k}}\biggr)
           -\ln\biggl(-\frac{A_{ki}\tau\tau_{\bar{\imath}j\bar{k}}+A_{ij}\tau_{j\bar{k}}\tau_{\bar{\imath}}}
                            {A_{jk}\tau_{\bar{\imath}j}\tau_{\bar{k}}}\biggr)\nn\\
\fl  &&    +\ln\biggl(-\frac{A_{ij}\tau_{\bar{\imath}k}\tau_{j}+A_{jk}\tau\tau_{\bar{\imath}jk}}{A_{ki}\tau_{\bar{\imath}j}\tau_{k}}\biggr)
           +\ln\biggl(-\frac{A_{ij}\tau_{i\bar{k}}\tau_{\bar{\jmath}}+A_{jk}\tau\tau_{i\bar{\jmath}\bar{k}}}
                            {A_{ki}\tau_{i\bar{\jmath}}\tau_{\bar{k}}}\biggr)\nn\\
\fl  &&    -\ln\biggl(-\frac{A_{ij}\tau\tau_{ij\bar{k}}+A_{jk}\tau_{i\bar{k}}\tau_{j}}{A_{ki}\tau_{j\bar{k}}\tau_{i}}\biggr)
           -\ln\biggl(-\frac{A_{ij}\tau\tau_{\bar{\imath}\bar{\jmath}k}+A_{jk}\tau_{\bar{\imath}k}\tau_{\bar{\jmath}}}
                            {A_{ki}\tau_{\bar{\jmath}k}\tau_{\bar{\imath}}}\biggr)\nn\\
\fl  &&    +\ln\biggl(-\frac{A_{jk}\tau_{i\bar{\jmath}}\tau_{k}+A_{ki}\tau\tau_{i\bar{\jmath}k}}{A_{ij}\tau_{\bar{\jmath}k}\tau_{i}}\biggr)
           +\ln\biggl(-\frac{A_{jk}\tau_{\bar{\imath}j}\tau_{\bar{k}}+A_{ki}\tau\tau_{\bar{\imath}j\bar{k}}}
                            {A_{ij}\tau_{j\bar{k}}\tau_{\bar{\imath}}}\biggr)\nn\\
\fl  &&    -\ln\biggl(-\frac{A_{jk}\tau\tau_{\bar{\imath}jk}+A_{ki}\tau_{\bar{\imath}j}\tau_{k}}{A_{ij}\tau_{\bar{\imath}k}\tau_{j}}\biggr)
           -\ln\biggl(-\frac{A_{jk}\tau\tau_{i\bar{\jmath}\bar{k}}+A_{ki}\tau_{i\bar{\jmath}}\tau_{\bar{k}}}
                            {A_{ij}\tau_{i\bar{k}}\tau_{\bar{\jmath}}}\biggr)\biggr\}\frac{1}{\tau}\nn\\
\fl & = & 0
\eea

Equation \eqref{12copies} is actually a 19-point equation existing on a cube as in Figure \ref{bigpicture}. It comprises the 12 shifted copies of \eqref{AKP} as illustrated in Figure \ref{6pictures}, where to each configuration of 6 points on an elementary cube correspond 2 copies of \eqref{AKP}.


\newcommand{\drawLinewithGBG}[2]
{
 \draw[gray!20,line width=3pt]  (#1) -- (#2);
 \draw[black,very thick] (#1) -- (#2);
}
\newcommand{\drawLinewithNBG}[2]
{
 \draw[black,very thick] (#1) -- (#2);
}
\newcommand{\drawArrow}[2]
{
 \draw[black,very thick,->] (#1) -- (#2);
}
\newcommand{\drawGridLine}[2]
{
 \draw[gray,very thin] (#1) -- (#2);
}
\newcommand{\drawDot}[1]
{
 \shade[ball color=black] (#1) circle (0.08);
}
\newcommand{\drawDiamond}[1]
{
 \shade[ball color=black] (#1) circle (0.08);
}
\newcommand{\drawGridLeft}
{
 \filldraw[fill=gray!20,fill opacity=0.5] (0,1) rectangle (0.6,0);
 \drawGridLine{0.3,1}{0.3,0}
 \drawGridLine{0,0.5}{0.6,0.5}
 \drawGridLine{0.8,0.375}{0.8,-0.625}
 \drawGridLine{0.5,-0.125}{1.1,-0.125}
}
\newcommand{\drawGridBottom}
{
 \filldraw[fill=gray!50,fill opacity=0.5] (0,0) rectangle (1.25,0.6);
 \filldraw[fill=gray!50,fill opacity=0.5] (-0.5,0.4) rectangle (0.75,1);
 \drawGridLine{0.625,0}{0.625,0.6}
 \drawGridLine{0,0.3}{1.25,0.3}
 \drawGridLine{0.125,0.4}{0.125,1}
 \drawGridLine{-0.5,0.7}{0.75,0.7}
}
\newcommand{\drawGridBack}
{
 \filldraw[fill=gray!20,fill opacity=0.5] (0.6,0.75) rectangle (1.6,1.75);
 \drawGridLine{1.1,0.75}{1.1,1.75}
 \drawGridLine{0.6,1.25}{1.6,1.25}
 \drawGridLine{0.8,0.375}{0.8,1.375}
 \drawGridLine{0.3,0.875}{1.3,0.875}
}
\newcommand{\drawGridRight}
{
 \filldraw[fill=gray!20,fill opacity=0.5] (1,-0.25) rectangle (1.6,-1.25);
 \drawGridLine{1.3,-0.25}{1.3,-1.25}
 \drawGridLine{1,-0.75}{1.6,-0.75}
}
\newcommand{\drawGridTop}
{
 \filldraw[fill=gray!50,fill opacity=0.5] (-1,0.8) rectangle (0.25,1.4);
 \drawGridLine{-0.375,0.8}{-0.375,1.4}
 \drawGridLine{-1,1.1}{0.25,1.1}
}
\newcommand{\drawGridFront}
{
 \filldraw[fill=gray!20,fill opacity=0.5] (0,0) rectangle (1,1);
 \drawGridLine{0.5,0}{0.5,1}
 \drawGridLine{0,0.5}{1,0.5}
}


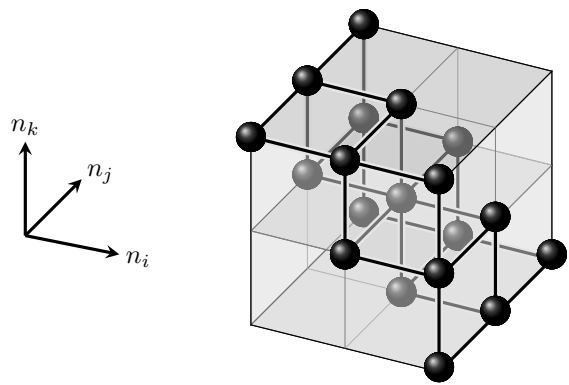
\begin{figure}[h]
\begin{center}
\begin{tikzpicture}[every node/.style={minimum size=1cm},scale=2.5,>=stealth]
 \drawArrow{-1.2,0.475}{-0.7,0.375}
 \drawArrow{-1.2,0.475}{-0.9,0.775}
 \drawArrow{-1.2,0.475}{-1.2,0.975}
 \node at (-0.6,0.35) {$n_{i}$};
 \node at (-0.8,0.8) {$n_{j}$};
 \node at (-1.2,1.05) {$n_{k}$};
\begin{scope}[every node/.append style={xslant=0,yslant=1},xslant=0,yslant=1]
  \drawGridLeft
\end{scope}
\begin{scope}[every node/.append style={xslant=1,yslant=-0.2},xslant=1,yslant=-0.2]
  \drawGridBottom
\end{scope}
 \begin{scope}[every node/.append style={xslant=0,yslant=-0.25},xslant=0,yslant=-0.25]
  \drawGridBack
  \drawLinewithGBG{0.3,0.875}{0.3,1.375}
  \drawLinewithGBG{0.3,0.875}{0.6,1.25}
  \drawLinewithGBG{0.5,0.5}{0.8,0.875}
  \drawLinewithGBG{0.6,0.75}{1.1,0.75}
  \drawLinewithGBG{0.6,1.25}{0.6,0.75}
  \drawLinewithGBG{0.6,1.25}{0.6,1.75}
  \drawLinewithGBG{0.8,0.375}{0.8,0.875}
  \drawLinewithGBG{0.8,0.375}{1.3,0.375}
  \drawLinewithGBG{0.8,0.875}{0.3,0.875}
  \drawLinewithGBG{0.8,0.875}{1.1,1.25}
  \drawLinewithGBG{1.1,0.75}{0.8,0.375}
  \drawLinewithGBG{1.1,0.75}{1.6,0.75}
  \drawLinewithGBG{1.1,1.25}{0.6,1.25}
  \drawLinewithGBG{1.1,1.25}{1.1,0.75}
  \drawLinewithGBG{1.3,0.875}{0.8,0.875}
  \drawLinewithGBG{0.8,1.375}{0.8,0.875}
\end{scope}
 \drawDot{1.1,0.975}
 \drawDot{1.1,0.475}
 \drawDot{0.8,0.175}
 \drawDot{0.6,1.1}
 \drawDot{0.6,0.6}
 \drawDot{0.3,0.8}
\begin{scope}[every node/.append style={xslant=0,yslant=-0.25},xslant=0,yslant=-0.25]
 \drawLinewithGBG{0.8,0.875}{0.5,0.5}
\end{scope}
\drawDiamond{0.8,0.675}
\begin{scope}[every node/.append style={xslant=0,yslant=1},xslant=0,yslant=1]
  \drawGridRight
\end{scope}
\begin{scope}[every node/.append style={xslant=1,yslant=-0.2},xslant=1,yslant=-0.2]
  \drawGridTop
\end{scope}
\begin{scope}[every node/.append style={xslant=0,yslant=-0.25},xslant=0,yslant=-0.25]
  \drawGridFront
  \drawLinewithNBG{1.6,0.75}{1.3,0.375}
  \drawLinewithGBG{1.3,0.375}{1.3,0.875}
  \drawLinewithNBG{0.3,1.375}{0,1}
  \drawLinewithGBG{0,1}{0.5,1}
  \drawLinewithGBG{0.5,1}{0.5,0.5}
  \drawLinewithGBG{0.8,1.375}{0.5,1}
  \drawLinewithGBG{0.5,1}{1,1}
  \drawLinewithGBG{1,1}{1,0.5}
  \drawLinewithGBG{1,0.5}{1.3,0.875}
  \drawLinewithNBG{0.6,1.75}{0.3,1.375}
  \drawLinewithGBG{0.3,1.375}{0.8,1.375}
  \drawLinewithNBG{1.3,0.375}{1,0}
  \drawLinewithGBG{1,0}{1,0.5}
  \drawLinewithGBG{1,0.5}{0.5,0.5}
\end{scope}
 \drawDot{1.6,0.375}
 \drawDot{1.3,0.575}
 \drawDot{1.3,0.075}
 \drawDot{1,0.775}
 \drawDot{1,0.275}
 \drawDot{1,-0.225}
 \drawDot{0.8,1.175}
 \drawDot{0.6,1.6}
 \drawDot{0.5,0.875}
 \drawDot{0.5,0.375}
 \drawDot{0,1}
 \drawDot{0.3,1.3}
\end{tikzpicture}
\caption{The 19-point equation.}
\label{bigpicture}
\end{center}
\end{figure}

\begin{figure}[h]
\begin{center}
\begin{tikzpicture}[every node/.style={minimum size=1cm},scale=2]
\begin{scope}[every node/.append style={xslant=0,yslant=1},xslant=0,yslant=1]
  \drawGridLeft
\end{scope}
\begin{scope}[every node/.append style={xslant=1,yslant=-0.2},xslant=1,yslant=-0.2]
  \drawGridBottom
\end{scope}
 \begin{scope}[every node/.append style={xslant=0,yslant=-0.25},xslant=0,yslant=-0.25]
  \drawGridBack
  \drawLinewithGBG{0.8,0.875}{1.1,1.25}
  \drawLinewithGBG{1.1,1.25}{1.1,0.75}
  \drawLinewithGBG{1.1,0.75}{1.6,0.75}
  \drawLinewithGBG{1.3,0.875}{0.8,0.875}
\end{scope}
 \drawDiamond{0.8,0.675}
 \drawDot{1.1,0.975}
 \drawDot{1.1,0.475}
\begin{scope}[every node/.append style={xslant=0,yslant=1},xslant=0,yslant=1]
  \drawGridRight
\end{scope}
\begin{scope}[every node/.append style={xslant=1,yslant=-0.2},xslant=1,yslant=-0.2]
  \drawGridTop
\end{scope}
\begin{scope}[every node/.append style={xslant=0,yslant=-0.25},xslant=0,yslant=-0.25]
  \drawGridFront
  \drawLinewithNBG{1.6,0.75}{1.3,0.375}
  \drawLinewithGBG{1.3,0.375}{1.3,0.875}
\end{scope}
 \drawDot{1.6,0.375}
 \drawDot{1.3,0.075}
 \drawDot{1.3,0.575}
\end{tikzpicture}
\begin{tikzpicture}[every node/.style={minimum size=1cm},scale=2]
\begin{scope}[every node/.append style={xslant=0,yslant=1},xslant=0,yslant=1]
  \drawGridLeft
\end{scope}
\begin{scope}[every node/.append style={xslant=1,yslant=-0.2},xslant=1,yslant=-0.2]
  \drawGridBottom
\end{scope}
 \begin{scope}[every node/.append style={xslant=0,yslant=-0.25},xslant=0,yslant=-0.25]
  \drawGridBack
  \drawLinewithGBG{0.5,0.5}{0.8,0.875}
  \drawLinewithGBG{0.8,0.875}{0.3,0.875}
  \drawLinewithGBG{0.3,0.875}{0.3,1.375}
\end{scope}
 \drawDiamond{0.8,0.675}
 \drawDot{0.3,0.8}
\begin{scope}[every node/.append style={xslant=0,yslant=1},xslant=0,yslant=1]
  \drawGridRight
\end{scope}
\begin{scope}[every node/.append style={xslant=1,yslant=-0.2},xslant=1,yslant=-0.2]
  \drawGridTop
\end{scope}
\begin{scope}[every node/.append style={xslant=0,yslant=-0.25},xslant=0,yslant=-0.25]
  \drawGridFront
  \drawLinewithNBG{0.3,1.375}{0,1}
  \drawLinewithGBG{0,1}{0.5,1}
  \drawLinewithGBG{0.5,1}{0.5,0.5}
\end{scope}
 \drawDot{0,1}
 \drawDot{0.3,1.3}
 \drawDot{0.5,0.875}
 \drawDot{0.5,0.375}
\end{tikzpicture}
\begin{tikzpicture}[every node/.style={minimum size=1cm},scale=2]
\begin{scope}[every node/.append style={xslant=0,yslant=1},xslant=0,yslant=1]
  \drawGridLeft
\end{scope}
\begin{scope}[every node/.append style={xslant=1,yslant=-0.2},xslant=1,yslant=-0.2]
  \drawGridBottom
\end{scope}
 \begin{scope}[every node/.append style={xslant=0,yslant=-0.25},xslant=0,yslant=-0.25]
  \drawGridBack
  \drawLinewithGBG{1.3,0.875}{0.8,0.875}
  \drawLinewithGBG{0.8,0.875}{0.8,1.375}
\end{scope}
 \drawDiamond{0.8,0.675}
\begin{scope}[every node/.append style={xslant=0,yslant=1},xslant=0,yslant=1]
  \drawGridRight
\end{scope}
\begin{scope}[every node/.append style={xslant=1,yslant=-0.2},xslant=1,yslant=-0.2]
  \drawGridTop
\end{scope}
\begin{scope}[every node/.append style={xslant=0,yslant=-0.25},xslant=0,yslant=-0.25]
  \drawGridFront
  \drawLinewithGBG{0.8,1.375}{0.5,1}
  \drawLinewithGBG{0.5,1}{1,1}
  \drawLinewithGBG{1,1}{1,0.5}
  \drawLinewithGBG{1,0.5}{1.3,0.875}
\end{scope}
 \drawDot{1.3,0.575}
 \drawDot{1,0.275}
 \drawDot{1,0.775}
 \drawDot{0.5,0.875}
 \drawDot{0.8,1.175}
\end{tikzpicture}
\begin{tikzpicture}[every node/.style={minimum size=1cm},scale=2]
\begin{scope}[every node/.append style={xslant=0,yslant=1},xslant=0,yslant=1]
  \drawGridLeft
\end{scope}
\begin{scope}[every node/.append style={xslant=1,yslant=-0.2},xslant=1,yslant=-0.2]
  \drawGridBottom
\end{scope}
 \begin{scope}[every node/.append style={xslant=0,yslant=-0.25},xslant=0,yslant=-0.25]
  \drawGridBack
  \drawLinewithGBG{0.3,0.875}{0.6,1.25}
  \drawLinewithGBG{0.6,1.25}{0.6,0.75}
  \drawLinewithGBG{0.6,0.75}{1.1,0.75}
  \drawLinewithGBG{1.1,0.75}{0.8,0.375}
  \drawLinewithGBG{0.8,0.375}{0.8,0.875}
  \drawLinewithGBG{0.8,0.875}{0.3,0.875}
\end{scope}
 \drawDiamond{0.8,0.675}
 \drawDot{0.3,0.8}
 \drawDot{0.6,1.1}
 \drawDot{0.6,0.6}
 \drawDot{0.8,0.175}
 \drawDot{1.1,0.475}
\begin{scope}[every node/.append style={xslant=0,yslant=1},xslant=0,yslant=1]
  \drawGridRight
\end{scope}
\begin{scope}[every node/.append style={xslant=1,yslant=-0.2},xslant=1,yslant=-0.2]
  \drawGridTop
\end{scope}
\begin{scope}[every node/.append style={xslant=0,yslant=-0.25},xslant=0,yslant=-0.25]
  \drawGridFront
\end{scope}
\end{tikzpicture}
\begin{tikzpicture}[every node/.style={minimum size=1cm},scale=2]
\begin{scope}[every node/.append style={xslant=0,yslant=1},xslant=0,yslant=1]
  \drawGridLeft
\end{scope}
\begin{scope}[every node/.append style={xslant=1,yslant=-0.2},xslant=1,yslant=-0.2]
  \drawGridBottom
\end{scope}
 \begin{scope}[every node/.append style={xslant=0,yslant=-0.25},xslant=0,yslant=-0.25]
  \drawGridBack
  \drawLinewithGBG{0.8,0.875}{1.1,1.25}
  \drawLinewithGBG{1.1,1.25}{0.6,1.25}
  \drawLinewithGBG{0.6,1.25}{0.6,1.75}
  \drawLinewithGBG{0.8,1.375}{0.8,0.875}
\end{scope}
 \drawDiamond{0.8,0.675}
 \drawDot{0.6,1.1}
 \drawDot{1.1,0.975}
\begin{scope}[every node/.append style={xslant=0,yslant=1},xslant=0,yslant=1]
  \drawGridRight
\end{scope}
\begin{scope}[every node/.append style={xslant=1,yslant=-0.2},xslant=1,yslant=-0.2]
  \drawGridTop
\end{scope}
\begin{scope}[every node/.append style={xslant=0,yslant=-0.25},xslant=0,yslant=-0.25]
  \drawGridFront
  \drawLinewithNBG{0.6,1.75}{0.3,1.375}
  \drawLinewithGBG{0.3,1.375}{0.8,1.375}
\end{scope}
 \drawDot{0.3,1.3}
 \drawDot{0.6,1.6}
 \drawDot{0.8,1.175}
\end{tikzpicture}
\begin{tikzpicture}[every node/.style={minimum size=1cm},scale=2]
\begin{scope}[every node/.append style={xslant=0,yslant=1},xslant=0,yslant=1]
  \drawGridLeft
\end{scope}
\begin{scope}[every node/.append style={xslant=1,yslant=-0.2},xslant=1,yslant=-0.2]
  \drawGridBottom
\end{scope}
 \begin{scope}[every node/.append style={xslant=0,yslant=-0.25},xslant=0,yslant=-0.25]
  \drawGridBack
  \drawLinewithGBG{0.5,0.5}{0.8,0.875}
  \drawLinewithGBG{0.8,0.875}{0.8,0.375}
  \drawLinewithGBG{0.8,0.375}{1.3,0.375}
\end{scope}
 \drawDiamond{0.8,0.675}
 \drawDot{0.8,0.175}
\begin{scope}[every node/.append style={xslant=0,yslant=1},xslant=0,yslant=1]
  \drawGridRight
\end{scope}
\begin{scope}[every node/.append style={xslant=1,yslant=-0.2},xslant=1,yslant=-0.2]
  \drawGridTop
\end{scope}
\begin{scope}[every node/.append style={xslant=0,yslant=-0.25},xslant=0,yslant=-0.25]
  \drawGridFront
  \drawLinewithNBG{1.3,0.375}{1,0}
  \drawLinewithGBG{1,0}{1,0.5}
  \drawLinewithGBG{1,0.5}{0.5,0.5}
\end{scope}
 \drawDot{1.3,0.075}
 \drawDot{1,-0.225}
 \drawDot{1,0.275}
 \drawDot{0.5,0.375}
\end{tikzpicture}
\caption{Copies of the 6-point equation.}
\label{6pictures}
\end{center}
\end{figure}
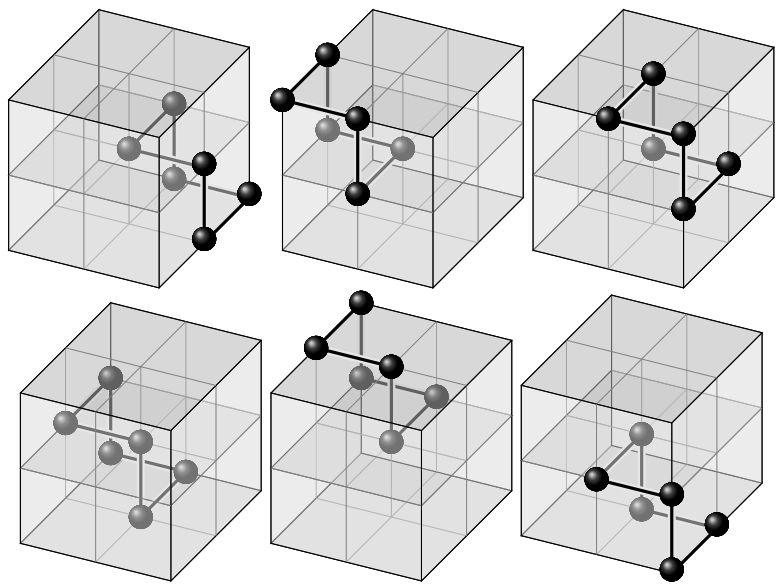

The main observation which allows the establishment of the multiform structure is that the Lagrangian 3-form defined in \eqref{L123} is a closed form on the solution space of the original bilinear equation \eqref{AKP}. In fact we have the following \emph{closure property}\\[10pt]
\textbf{Proposition:} \emph{The Lagrangian defined by \eqref{L123} satisfies the following closure relation on solutions to the equation \eqref{AKP} when embedded in a 4-dimensional lattice.
\be\label{3dclos}
 \Dl_{l}\cL_{ijk}-\Dl_{i}\cL_{jkl}+\Dl_{j}\cL_{kli}-\Dl_{k}\cL_{lij} = 0,
\ee
where the difference operator $\Dl_{i}$ acts on functions $f$ of $\tau=\tau(n_{i},n_{j},n_{k},n_{l})$ by the formula $\Dl_{i}f(\tau)=f(\tau_{i})-f(\tau)$, and on a function $g$ of $\tau$ and its shifts by the formula $\Dl_{i}g(\tau,\tau_{j},\tau_{k},\tau_{l})=g(\tau_{i},\tau_{ij},\tau_{ik},\tau_{il})-g(\tau,\tau_{j},\tau_{k},\tau_{l})$.}\\[10pt]
\textit{Proof:} By explicit computation. The closure relation \eqref{3dclos} holds on solutions of the original equation, so we need to make use of \eqref{AKP} and its shifted versions. If we add in a fourth lattice direction, we get the equations

\bse
\bea
A_{jk}\tau_{i}\tau_{jk}+A_{ki}\tau_{j}\tau_{ki}+A_{ij}\tau_{k}\tau_{ij}=0&&\label{AKP1},\\
A_{kl}\tau_{j}\tau_{kl}-A_{jl}\tau_{k}\tau_{jl}+A_{jk}\tau_{l}\tau_{jk}=0&&\label{AKP2},
\eea
\bea
A_{li}\tau_{k}\tau_{li}-A_{ki}\tau_{l}\tau_{ki}+A_{kl}\tau_{i}\tau_{kl}=0&&\label{AKP3},\\
A_{ij}\tau_{l}\tau_{ij}+A_{jl}\tau_{i}\tau_{jl}+A_{li}\tau_{j}\tau_{li}=0&&\label{AKP4}.
\eea
When shifted, equations \eqref{AKP1} through \eqref{AKP4} become
\bea
A_{jk}\tau_{li}\tau_{jkl}+A_{ki}\tau_{jl}\tau_{kli}+A_{ij}\tau_{kl}\tau_{lij}=0&&\label{AKP5},\\
A_{kl}\tau_{ij}\tau_{kli}-A_{jl}\tau_{ki}\tau_{lij}+A_{jk}\tau_{li}\tau_{ijk}=0&&\label{AKP6},\\
A_{li}\tau_{jk}\tau_{lij}-A_{ki}\tau_{jl}\tau_{ijk}+A_{kl}\tau_{ij}\tau_{jkl}=0&&\label{AKP7},\\
A_{ij}\tau_{kl}\tau_{ijk}+A_{jl}\tau_{ki}\tau_{jkl}+A_{li}\tau_{jk}\tau_{kli}=0&&\label{AKP8}.
\eea
\ese
We also need the following two key identities for the dilogarithm function
\bse
\bea
  \Li_2(x)+\Li_2(y) & = & \Li_2(xy)-\Li_2\biggl(\frac{x-xy}{x-1}\biggr)-\Li_2\biggl(\frac{y-xy}{y-1}\biggr)\nn\\
                       && -\frac{1}{2}\biggl(\ln\biggl(\frac{x-1}{y-1}\biggr)\biggr)^2,\label{5ptn}\\
  \Li_2(x)+\Li_2\biggl(\frac{1}{x}\biggr) & = & -\frac{1}{2}\bigl(\ln(-x)\bigr)^2-\frac{\pi^2}{6},\label{flip}
\eea
\ese
The latter equation holds for all $x$, a proof of which can be found in \cite{Lewin}, where many dilogarithm identities are collected. Equation \eqref{5ptn} is a combination of other such identities from \cite{Lewin}, and it can be proved by simple differentiation. It is valid up to imaginary terms which can be chosen to cancel out in the course of the closure relation computation.
We will split the computation into two parts, considering the dilogarithm terms separately. Let 
\be\label{Gamma}
 \Gamma = \Dl_{l}\cL_{ijk}-\Dl_{i}\cL_{jkl}+\Dl_{j}\cL_{kli}-\Dl_{k}\cL_{lij}
\ee
with $\cL_{ijk}$ given by \eqref{L123} and let $\Gamma = \Gamma_1 + \Gamma_2$, where $\Gamma_1$ is the part of $\Gamma$ omitting dilogarithm terms from the Lagrangian, and $\Gamma_2$ consists of only the dilogarithm terms. We have

\bea\label{Gamma1}
\fl \Gamma_1 & = & \frac{1}{2}\bigl((\ln(\tau_{ijk})^2-(\ln(\tau_{jkl}))^2+(\ln(\tau_{kli}))^2-(\ln(\tau_{lij}))^2\nn\\
\fl      && \;\;\;\;\;                        +(\ln(\tau_{i}))^2-(\ln(\tau_{j}))^2+(\ln(\tau_{k}))^2-(\ln(\tau_{l}))^2\bigr)\nn\\
\fl      && -\ln(\tau_{ijk})\ln(\tau_{kli})+\ln(\tau_{jkl})\ln(\tau_{lij})-\ln(\tau_{i})\ln(\tau_{k})+\ln(\tau_{j})\ln(\tau_{l})\nn\\
\fl      && +\ln(\tau_{ijk})\ln\biggl(-\frac{A_{ij}A_{jk}\tau_{kl}\tau_{li}}{A_{jl}A_{ki}\tau_{jl}\tau_{ki}}\biggr)
            +\ln(\tau_{jkl})\ln\biggl(\frac{A_{jl}A_{ki}\tau_{jl}\tau_{ki}}{A_{jk}A_{kl}\tau_{ij}\tau_{li}}\biggr)\nn\\
\fl      && +\ln(\tau_{kli})\ln\biggl(-\frac{A_{kl}A_{li}\tau_{ij}\tau_{jk}}{A_{jl}A_{ki}\tau_{jl}\tau_{ki}}\biggr)
            +\ln(\tau_{lij})\ln\biggl(\frac{A_{jl}A_{ki}\tau_{jl}\tau_{ki}}{A_{ij}A_{li}\tau_{jk}\tau_{kl}}\biggr)\nn\\
\fl      && +\ln(\tau_{i})\ln\biggl(\frac{A_{jk}A_{kl}\tau_{jk}\tau_{kl}}{A_{jl}A_{ki}\tau_{jl}\tau_{ki}}\biggr)
            +\ln(\tau_{j})\ln\biggl(-\frac{A_{jl}A_{ki}\tau_{jl}\tau_{ki}}{A_{kl}A_{li}\tau_{kl}\tau_{li}}\biggr)\nn\\
\fl      && +\ln(\tau_{k})\ln\biggl(\frac{A_{ij}A_{li}\tau_{ij}\tau_{li}}{A_{jl}A_{ki}\tau_{jl}\tau_{ki}}\biggr)
            +\ln(\tau_{l})\ln\biggl(-\frac{A_{jl}A_{ki}\tau_{jl}\tau_{ki}}{A_{ij}A_{jk}\tau_{ij}\tau_{jk}}\biggr)\nn\\
\fl      && +\ln(\tau_{ij})\ln\biggl(-\frac{A_{li}}{A_{jk}}\biggr)+\ln(\tau_{jk})\ln\biggl(-\frac{A_{kl}}{A_{ij}}\biggr)
            +\ln(\tau_{kl})\ln\biggl(-\frac{A_{jk}}{A_{li}}\biggr)\nn\\
\fl      && +\ln(\tau_{li})\ln\biggl(-\frac{A_{ij}}{A_{kl}}\biggr)
            +\ln\biggl(\frac{\tau_{jl}}{\tau_{ki}}\biggr)\ln\biggl(\frac{A_{ij}A_{jk}A_{kl}A_{li}\tau_{ij}\tau_{jk}\tau_{kl}\tau_{li}}
                                                                        {A_{jl}^2A_{ki}^2\tau_{jl}^2\tau_{ki}^2}\biggr).
\eea

Now we consider the dilogarithm terms. The dilogarithm terms from $\Gamma$ are

\setlength{\unitlength}{1cm}
\begin{picture}(10,10)(0,0)
 \put(0,10){\parbox[t]{6cm}{\bea\label{eqbox}
\fl \Gamma_2 & = & -\Li_{2}\biggl(-\frac{A_{ij}\tau_{kl}\tau_{lij}}{A_{ki}\tau_{jl}\tau_{kli}}\biggr)
                 \;+\Li_{2}\biggl(-\frac{A_{ij}\tau_{k}\tau_{ij}}{A_{ki}\tau_{j}\tau_{ki}}\biggr)
                 \;\;\;\;-\Li_{2}\biggl(\frac{A_{kl}\tau_{ij}\tau_{jkl}}{A_{ki}\tau_{jl}\tau_{ijk}}\biggr)\nn\\
\fl           && \nn\\
\fl           && +\Li_{2}\biggl(\frac{A_{jk}\tau_{li}\tau_{ijk}}{A_{jl}\tau_{ki}\tau_{lij}}\biggr)
                 \;\;\;\;-\Li_{2}\biggl(\frac{A_{jk}\tau_{l}\tau_{jk}}{A_{jl}\tau_{k}\tau_{jl}}\biggr)
                 \;\;\;\;\;\;\;+\Li_{2}\biggl(\frac{A_{kl}\tau_{i}\tau_{kl}}{A_{ki}\tau_{l}\tau_{ki}}\biggr)\nn\\
\fl           && \nn\\
\fl           && +\Li_{2}\biggl(-\frac{A_{li}\tau_{jk}\tau_{kli}}{A_{jl}\tau_{ki}\tau_{jkl}}\biggr)
                 \;-\Li_{2}\biggl(-\frac{A_{li}\tau_{j}\tau_{li}}{A_{jl}\tau_{i}\tau_{jl}}\biggr)
                 \;\;\;\;+\Li_{2}\biggl(-\frac{A_{kl}\tau_{ij}\tau_{kli}}{A_{jk}\tau_{li}\tau_{ijk}}\biggr)\nn\\
\fl           && \nn\\
\fl           && -\Li_{2}\biggl(-\frac{A_{jk}\tau_{li}\tau_{jkl}}{A_{ij}\tau_{kl}\tau_{lij}}\biggr)
                 \;+\Li_{2}\biggl(-\frac{A_{jk}\tau_{i}\tau_{jk}}{A_{ij}\tau_{k}\tau_{ij}}\biggr)
                 \;\;\;\;-\Li_{2}\biggl(-\frac{A_{kl}\tau_{j}\tau_{kl}}{A_{jk}\tau_{l}\tau_{jk}}\biggr)\nn\\
\fl           && \nn\\
\fl           && -\Li_{2}\biggl(-\frac{A_{li}\tau_{jk}\tau_{lij}}{A_{kl}\tau_{ij}\tau_{jkl}}\biggr)
                 \;+\Li_{2}\biggl(-\frac{A_{li}\tau_{k}\tau_{li}}{A_{kl}\tau_{i}\tau_{kl}}\biggr)
                 \;\;\;\;-\Li_{2}\biggl(-\frac{A_{ki}\tau_{jl}\tau_{kli}}{A_{jk}\tau_{li}\tau_{jkl}}\biggr)\nn\\
\fl           && \nn\\
\fl           && +\Li_{2}\biggl(-\frac{A_{ij}\tau_{kl}\tau_{ijk}}{A_{li}\tau_{jk}\tau_{kli}}\biggr)
                 \;-\Li_{2}\biggl(-\frac{A_{ij}\tau_{l}\tau_{ij}}{A_{li}\tau_{j}\tau_{li}}\biggr)
                 \;\;\;\;+\Li_{2}\biggl(-\frac{A_{ki}\tau_{j}\tau_{ki}}{A_{jk}\tau_{i}\tau_{jk}}\biggr)\nn\\
\fl           && \nn\\
\fl           && +\Li_{2}\biggl(\frac{A_{jl}\tau_{ki}\tau_{lij}}{A_{kl}\tau_{ij}\tau_{kli}}\biggr)               
                 \;\;\;\;-\Li_{2}\biggl(\frac{A_{jl}\tau_{k}\tau_{jl}}{A_{kl}\tau_{j}\tau_{kl}}\biggr)
                 \;\;\;\;\;\;\;+\Li_{2}\biggl(-\frac{A_{jl}\tau_{ki}\tau_{jkl}}{A_{ij}\tau_{kl}\tau_{ijk}}\biggr)\nn\\
\fl           && \nn\\
\fl           && -\Li_{2}\biggl(\frac{A_{ki}\tau_{jl}\tau_{ijk}}{A_{li}\tau_{jk}\tau_{lij}}\biggr)               
                 \;\;\;\;+\Li_{2}\biggl(\frac{A_{ki}\tau_{l}\tau_{ki}}{A_{li}\tau_{k}\tau_{li}}\biggr)
                 \;\;\;\;\;\;\;-\Li_{2}\biggl(-\frac{A_{jl}\tau_{i}\tau_{jl}}{A_{ij}\tau_{l}\tau_{ij}}\biggr)\nn\\
\fl           && \nn\\
\fl           && 
\eea}}
 \put(0.9,8.9){\dashbox{0.1}(3.3,0.9)}
 \put(0.9,5.3){\dashbox{0.1}(3.3,0.95)}
 \put(0.9,4.1){\dashbox{0.1}(3.3,0.95)}
 \put(0.9,0.55){\dashbox{0.1}(3.1,0.9)}
 \put(4.2,7.7){\dashbox{0.1}(2.9,0.9)}
 \put(4.2,6.5){\dashbox{0.1}(3.1,0.9)}
 \put(4.2,2.95){\dashbox{0.1}(3.1,0.9)}
 \put(4.2,1.75){\dashbox{0.1}(2.9,0.9)}
 \put(7.6,8.9){\dashbox{0.1}(3.1,0.9)}
 \put(7.6,5.3){\dashbox{0.1}(3.1,0.95)}
 \put(7.6,4.1){\dashbox{0.1}(3.4,0.95)}
 \put(7.6,0.6){\dashbox{0.1}(3.1,0.9)}
 \put(0.8,8.8){\framebox(6.5,1.1)}
 \put(0.8,7.6){\framebox(6.4,1.1)}
 \put(0.8,6.4){\framebox(6.6,1.1)}
 \put(0.8,5.2){\framebox(6.5,1.1)}
 \put(0.8,4){\framebox(6.5,1.1)}
 \put(0.8,2.8){\framebox(6.6,1.1)}
 \put(0.8,1.6){\framebox(6.5,1.1)}
 \put(0.8,0.4){\framebox(6.3,1.1)}
 \put(7.5,7.6){\framebox(3.3,2.3)}
 \put(7.5,5.2){\framebox(3.5,2.3)}
 \put(7.5,2.8){\framebox(3.6,2.3)}
 \put(7.5,0.4){\framebox(3.5,2.3)}
\end{picture}

Using \eqref{flip} on the terms in the dotted boxes, followed by \eqref{5ptn} on the terms in the solid boxes gives a large expression which we reproduce in the Appendix and show to be equal to $-\Gamma_1$, verifying the closure relation. $\blacksquare$

\paragraph{}
The establishment of the closure property enables us to propose a novel variational principle for the multidimensionally consistent system of bilinear KP equations, along the same line as in \cite{closure}. Choosing a 3-dimensional hypersurface $\sigma$ within a multidimensional lattice of dimension higher than 3, consisting of a connected configuration of elementary cubes $\sigma_{ijk}$, we can define an action $S$ on this hypersurface by summing the contributions $\cL_{ijk}$ from each of the cubes as follows
\be
 S[\tau;\sigma] = \sum_{\sigma_{ijk}\in\sigma} \cL_{ijk},
\ee
taking into consideration the orientation of each elementary cube contributing to the surface. The antisymmetry of $\cL_{ijk}$ guarantees that there is no ambiguity in how each discrete Lagrangian 3-form will contribute to the action \footnote{Note that we do not use the abstract notation of difference forms as proposed in \cite{Hydon}, as we prefer to work with the explicit expressions for the integrated forms as described here, which allows for a direct verification of the main result.}. Furthermore, the closure relation \eqref{3dclos} allows us to impose the independence of the action on local variations of the surface away from any boundary that the surface $\sigma$ may possess. Thus, whilst keeping the boundary fixed we may locally deform $\sigma$ in any way we choose, allowing us in particular to render it locally flat away from the boundary, such that we can specify a 3-dimensional hypersurface described in terms of three local coordinates $n_{i},n_{j},n_{k}$. There we can then apply the usual variational principle, taking the variational derivative with respect to $\tau$, leading to the Euler-Lagrange equations \eqref{12copies}. These equations of the motion are a consequence of the Hirota-Miwa equation \eqref{AKP}, as are the closure relations that guarantee the surface independence of the action under local deformations. This interlinked scheme of variations with respect to the dependent variables as well as to the geometry of the independent variables is what constitutes the Lagrangian multiform structure of the lattice KP system.

\section{Discussion}
We have shown that the ideas of \cite{closure} for 2-dimensional integrable (in the sense of multidimensional consistency) equations are also applicable to the 3-dimensional example of the bilinear discrete KP equation. There are several remarks we would like to make at this point. 

First, one has to qualify what it means for a Lagrangian to be associated with a given equation, since as we have noted earlier the Euler-Lagrange equations rather than yielding the original bilinear KP equation only yield a derived equation comprising a combination of various copies of the original equation. Nevertheless we have taken the point of view that since the canonical variable is the $\tau$-function we consider this Lagrangian structure to be associated with the bilinear KP equation.

Second, the closure relation which is central to the Lagrangian multiform structure relies on the bilinear KP equation rather than on the Euler-Lagrange equations. It is not clear at this stage to what extent the closure property remains to be verified on all solutions of the Euler-Lagrange equations or only on a subvariety of solutions that obey the multidimensional systems of bilinear equations. 

Third, we consider the Lagrangian multiform structure as a hallmark of multidimensional consistency on the level of the variational principle. As such, it is as much a principle that selects ``admissable Lagrangians'' as well as field configurations obeying the variational equations. It would be a challenge to see whether this principle can be used as a criterion to classify the admissable Lagrangians to which it can be applied, which then necessarily would coincide with the integrable cases.

As far as KP-type systems are concerned, in some recent works in combinatorics 3-dimensional 6-point recurrence schemes have been studied from the point of view of the geometry of the octahedral lattice, cf e.g.\cite{Speyer,Henriques}. A classification of multidimensionally consistent 6-point equations has recently been done in \cite{ABS_oct}, but this does not seem to yield any novel lattice equations (e.g. in comparison with the list in \cite{Frank1990}). It would be of interest to see whether Lagrangian multiform structures can be established for all those equations, and whether these structures can be adapted to the octahedral lattice picture. Alternatively one can consider 3-dimensional lattice equations of BKP type, i.e. equations of the form 
\be
 Q(\tau,\tau_{i},\tau_{j},\tau_{k},\tau_{ij},\tau_{jk},\tau_{ki},\tau_{ijk})=0
\ee
but so far Lagrangian structures for such equations remain to be established. 

\appendix
\section{Closure relation computation}
After using the dilogarithm identities on the boxed terms of equation \eqref{eqbox} as described above, we obtain the following expression. Here we have also made use of the equations \eqref{AKP1} through \eqref{AKP8}.

\setlength{\unitlength}{1cm}
\begin{picture}(10,20)(0,0)
 \put(0,19.75){\parbox[t]{6cm}{\bea
\fl \Gamma_2 & = & +\Li_{2}\biggl(\frac{\tau_{k}\tau_{ij}\tau_{jl}\tau_{kli}}{\tau_{j}\tau_{kl}\tau_{ki}\tau_{lij}}\biggr)
                 \;-\Li_{2}\biggl(\frac{\tau_{i}\tau_{jk}\tau_{jl}\tau_{kli}}{\tau_{j}\tau_{li}\tau_{ki}\tau_{jkl}}\biggr)
                 \;-\Li_{2}\biggl(\frac{\tau_{k}\tau_{ij}\tau_{li}\tau_{jkl}}{\tau_{i}\tau_{jk}\tau_{kl}\tau_{lij}}\biggr)\nn\\
\fl           && +\Li_{2}\biggl(\frac{\tau_{k}\tau_{li}\tau_{jl}\tau_{ijk}}{\tau_{l}\tau_{jk}\tau_{ki}\tau_{lij}}\biggr)
                 \;-\Li_{2}\biggl(\frac{\tau_{j}\tau_{kl}\tau_{li}\tau_{ijk}}{\tau_{l}\tau_{ij}\tau_{jk}\tau_{kli}}\biggr)
                 \;-\Li_{2}\biggl(\frac{\tau_{k}\tau_{ij}\tau_{jl}\tau_{kli}}{\tau_{j}\tau_{kl}\tau_{ki}\tau_{lij}}\biggr)\nn\\
\fl           && +\Li_{2}\biggl(\frac{\tau_{i}\tau_{kl}\tau_{jl}\tau_{ijk}}{\tau_{l}\tau_{ij}\tau_{ki}\tau_{jkl}}\biggr)
                 \;-\Li_{2}\biggl(\frac{\tau_{k}\tau_{li}\tau_{jl}\tau_{ijk}}{\tau_{l}\tau_{jk}\tau_{ki}\tau_{lij}}\biggr)
                 \;-\Li_{2}\biggl(\frac{\tau_{i}\tau_{jk}\tau_{kl}\tau_{lij}}{\tau_{k}\tau_{ij}\tau_{li}\tau_{jkl}}\biggr)\nn\\
\fl           && +\Li_{2}\biggl(\frac{\tau_{i}\tau_{jk}\tau_{jl}\tau_{kli}}{\tau_{j}\tau_{li}\tau_{ki}\tau_{jkl}}\biggr)
                 \;-\Li_{2}\biggl(\frac{\tau_{l}\tau_{ij}\tau_{jk}\tau_{kli}}{\tau_{j}\tau_{kl}\tau_{li}\tau_{ijk}}\biggr)
                 \;-\Li_{2}\biggl(\frac{\tau_{i}\tau_{kl}\tau_{jl}\tau_{ijk}}{\tau_{l}\tau_{ij}\tau_{ki}\tau_{jkl}}\biggr)\nn\\      
\fl           && +\Li_{2}\biggl(\frac{\tau_{i}\tau_{jk}\tau_{kl}\tau_{lij}}{\tau_{k}\tau_{ij}\tau_{li}\tau_{jkl}}\biggr)
                 \;-\Li_{2}\biggl(\frac{\tau_{j}\tau_{kl}\tau_{ki}\tau_{lij}}{\tau_{k}\tau_{ij}\tau_{jl}\tau_{kli}}\biggr)
                 \;-\Li_{2}\biggl(\frac{\tau_{i}\tau_{jk}\tau_{jl}\tau_{kli}}{\tau_{j}\tau_{li}\tau_{ki}\tau_{jkl}}\biggr)\nn\\
\fl           && +\Li_{2}\biggl(\frac{\tau_{l}\tau_{ij}\tau_{jk}\tau_{kli}}{\tau_{j}\tau_{kl}\tau_{li}\tau_{ijk}}\biggr)
                 \;-\Li_{2}\biggl(\frac{\tau_{k}\tau_{ij}\tau_{jl}\tau_{kli}}{\tau_{j}\tau_{kl}\tau_{ki}\tau_{lij}}\biggr)
                 \;-\Li_{2}\biggl(\frac{\tau_{l}\tau_{jk}\tau_{ki}\tau_{lij}}{\tau_{k}\tau_{li}\tau_{jl}\tau_{ijk}}\biggr)\nn\\
\fl           && +\Li_{2}\biggl(\frac{\tau_{k}\tau_{ij}\tau_{li}\tau_{jkl}}{\tau_{i}\tau_{jk}\tau_{kl}\tau_{lij}}\biggr)
                 \;-\Li_{2}\biggl(\frac{\tau_{l}\tau_{ij}\tau_{ki}\tau_{jkl}}{\tau_{i}\tau_{kl}\tau_{jl}\tau_{ijk}}\biggr)
                 \;-\Li_{2}\biggl(\frac{\tau_{k}\tau_{li}\tau_{jl}\tau_{ijk}}{\tau_{l}\tau_{jk}\tau_{ki}\tau_{lij}}\biggr)\nn\\
\fl           && +\Li_{2}\biggl(\frac{\tau_{j}\tau_{kl}\tau_{li}\tau_{ijk}}{\tau_{l}\tau_{ij}\tau_{jk}\tau_{kli}}\biggr)
                 \;-\Li_{2}\biggl(\frac{\tau_{i}\tau_{kl}\tau_{jl}\tau_{ijk}}{\tau_{l}\tau_{ij}\tau_{ki}\tau_{jkl}}\biggr)
                 \;-\Li_{2}\biggl(\frac{\tau_{j}\tau_{li}\tau_{ki}\tau_{jkl}}{\tau_{i}\tau_{jk}\tau_{jl}\tau_{kli}}\biggr)\nn\\
\fl           && +\Li_{2}\biggl(\frac{\tau_{j}\tau_{li}\tau_{ki}\tau_{jkl}}{\tau_{i}\tau_{jk}\tau_{jl}\tau_{kli}}\biggr)
                 \;-\Li_{2}\biggl(\frac{\tau_{k}\tau_{ij}\tau_{li}\tau_{jkl}}{\tau_{i}\tau_{jk}\tau_{kl}\tau_{lij}}\biggr)
                 \;-\Li_{2}\biggl(\frac{\tau_{j}\tau_{kl}\tau_{ki}\tau_{lij}}{\tau_{k}\tau_{ij}\tau_{jl}\tau_{kli}}\biggr)\nn\\
\fl           && +\Li_{2}\biggl(\frac{\tau_{j}\tau_{kl}\tau_{ki}\tau_{lij}}{\tau_{k}\tau_{ij}\tau_{jl}\tau_{kli}}\biggr)
                 \;-\Li_{2}\biggl(\frac{\tau_{l}\tau_{jk}\tau_{ki}\tau_{lij}}{\tau_{k}\tau_{li}\tau_{jl}\tau_{ijk}}\biggr)
                 \;-\Li_{2}\biggl(\frac{\tau_{j}\tau_{kl}\tau_{li}\tau_{ijk}}{\tau_{l}\tau_{ij}\tau_{jk}\tau_{kli}}\biggr)\nn\\
\fl           && +\Li_{2}\biggl(\frac{\tau_{l}\tau_{jk}\tau_{ki}\tau_{lij}}{\tau_{k}\tau_{li}\tau_{jl}\tau_{ijk}}\biggr)
                 \;-\Li_{2}\biggl(\frac{\tau_{i}\tau_{jk}\tau_{kl}\tau_{lij}}{\tau_{k}\tau_{ij}\tau_{li}\tau_{jkl}}\biggr)
                 \;-\Li_{2}\biggl(\frac{\tau_{l}\tau_{ij}\tau_{ki}\tau_{jkl}}{\tau_{i}\tau_{kl}\tau_{jl}\tau_{ijk}}\biggr)\nn\\
\fl           && +\Li_{2}\biggl(\frac{\tau_{l}\tau_{ij}\tau_{ki}\tau_{jkl}}{\tau_{i}\tau_{kl}\tau_{jl}\tau_{ijk}}\biggr)
                 \;-\Li_{2}\biggl(\frac{\tau_{j}\tau_{li}\tau_{ki}\tau_{jkl}}{\tau_{i}\tau_{jk}\tau_{jl}\tau_{kli}}\biggr)
                 \;-\Li_{2}\biggl(\frac{\tau_{l}\tau_{ij}\tau_{jk}\tau_{kli}}{\tau_{j}\tau_{kl}\tau_{li}\tau_{ijk}}\biggr)\nn\\
\fl           && +\frac{1}{2}\biggl(\ln\biggl(\frac{A_{ki}\tau_{jl}\tau_{kli}}{A_{ij}\tau_{kl}\tau_{lij}}\biggr)\biggr)^2
                 +\frac{1}{2}\biggl(\ln\biggl(-\frac{A_{jl}\tau_{k}\tau_{jl}}{A_{jk}\tau_{l}\tau_{jk}}\biggr)\biggr)^2
                 +\frac{1}{2}\biggl(\ln\biggl(-\frac{A_{ki}\tau_{jl}\tau_{ijk}}{A_{kl}\tau_{ij}\tau_{jkl}}\biggr)\biggr)^2\nn\\
\fl           && +\frac{1}{2}\biggl(\ln\biggl(\frac{A_{jl}\tau_{i}\tau_{jl}}{A_{li}\tau_{j}\tau_{li}}\biggr)\biggr)^2
                 +\frac{1}{2}\biggl(\ln\biggl(\frac{A_{ij}\tau_{kl}\tau_{lij}}{A_{jk}\tau_{li}\tau_{jkl}}\biggr)\biggr)^2
                 +\frac{1}{2}\biggl(\ln\biggl(\frac{A_{jk}\tau_{l}\tau_{jk}}{A_{kl}\tau_{j}\tau_{kl}}\biggr)\biggr)^2\nn\\
\fl           && +\frac{1}{2}\biggl(\ln\biggl(\frac{A_{kl}\tau_{ij}\tau_{jkl}}{A_{li}\tau_{jk}\tau_{lij}}\biggr)\biggr)^2
                 +\frac{1}{2}\biggl(\ln\biggl(\frac{A_{li}\tau_{j}\tau_{li}}{A_{ij}\tau_{l}\tau_{ij}}\biggr)\biggr)^2
                 +\frac{1}{2}\biggl(\ln\biggl(\frac{A_{jk}\tau_{li}\tau_{jkl}}{A_{ki}\tau_{jl}\tau_{kli}}\biggr)\biggr)^2\nn\\             
\fl           && +\frac{1}{2}\biggl(\ln\biggl(-\frac{A_{kl}\tau_{j}\tau_{kl}}{A_{jl}\tau_{k}\tau_{jl}}\biggr)\biggr)^2
                 +\frac{1}{2}\biggl(\ln\biggl(-\frac{A_{li}\tau_{jk}\tau_{lij}}{A_{ki}\tau_{jl}\tau_{ijk}}\biggr)\biggr)^2
                 +\frac{1}{2}\biggl(\ln\biggl(\frac{A_{ij}\tau_{l}\tau_{ij}}{A_{jl}\tau_{i}\tau_{jl}}\biggr)\biggr)^2\nn\\
\fl           && -\frac{1}{2}\biggl(\ln\biggl(\frac{A_{ki}\tau_{j}\tau_{li}\tau_{ki}\tau_{jkl}}
                                                   {A_{ij}\tau_{i}\tau_{jk}\tau_{kl}\tau_{lij}}\biggr)\biggr)^2
                 -\frac{1}{2}\biggl(\ln\biggl(-\frac{A_{jk}\tau_{l}\tau_{ij}\tau_{jk}\tau_{kli}}
                                                    {A_{jl}\tau_{j}\tau_{kl}\tau_{ki}\tau_{lij}}\biggr)\biggr)^2\nn\\
\fl           && -\frac{1}{2}\biggl(\ln\biggl(-\frac{A_{ki}\tau_{l}\tau_{jk}\tau_{ki}\tau_{lij}}
                                                    {A_{kl}\tau_{k}\tau_{ij}\tau_{li}\tau_{jkl}}\biggr)\biggr)^2
                 -\frac{1}{2}\biggl(\ln\biggl(\frac{A_{li}\tau_{j}\tau_{kl}\tau_{li}\tau_{ijk}}
                                                   {A_{jl}\tau_{l}\tau_{ij}\tau_{ki}\tau_{jkl}}\biggr)\biggr)^2\nn\\
\fl           && -\frac{1}{2}\biggl(\ln\biggl(\frac{A_{ij}\tau_{k}\tau_{ij}\tau_{jl}\tau_{kli}}
                                                   {A_{jk}\tau_{j}\tau_{li}\tau_{ki}\tau_{jkl}}\biggr)\biggr)^2
                 -\frac{1}{2}\biggl(\ln\biggl(\frac{A_{kl}\tau_{j}\tau_{kl}\tau_{ki}\tau_{lij}}
                                                   {A_{jk}\tau_{k}\tau_{li}\tau_{jl}\tau_{ijk}}\biggr)\biggr)^2
 \eea}}
 \put(7.4,16.6){\dashbox{0.1}(3.3,0.9)}
 \put(4,15.6){\dashbox{0.1}(3.3,0.9)}
 \put(4,13.6){\dashbox{0.1}(3.3,1)}
 \put(4,11.65){\dashbox{0.1}(3.3,0.9)}
 \put(7.4,12.6){\dashbox{0.1}(3.3,1)}
 \put(7.4,11.65){\dashbox{0.1}(3.3,0.9)}
\end{picture}

\bea                   
\fl           && \;\;\;\;\;-\frac{1}{2}\biggl(\ln\biggl(\frac{A_{kl}\tau_{i}\tau_{kl}\tau_{jl}\tau_{ijk}}
                                                   {A_{li}\tau_{l}\tau_{jk}\tau_{ki}\tau_{lij}}\biggr)\biggr)^2
                 -\frac{1}{2}\biggl(\ln\biggl(\frac{A_{ij}\tau_{l}\tau_{ij}\tau_{ki}\tau_{jkl}}
                                                   {A_{li}\tau_{i}\tau_{jk}\tau_{jl}\tau_{kli}}\biggr)\biggr)^2\nn\\
\fl           && \;\;\;\;\;-\frac{1}{2}\biggl(\ln\biggl(\frac{A_{jk}\tau_{i}\tau_{jk}\tau_{kl}\tau_{lij}}
                                                   {A_{ki}\tau_{k}\tau_{ij}\tau_{jl}\tau_{kli}}\biggr)\biggr)^2
                 -\frac{1}{2}\biggl(\ln\biggl(-\frac{A_{jl}\tau_{k}\tau_{li}\tau_{jl}\tau_{ijk}}
                                                    {A_{kl}\tau_{l}\tau_{ij}\tau_{jk}\tau_{kli}}\biggr)\biggr)^2\nn\\
\fl           && \;\;\;\;\;-\frac{1}{2}\biggl(\ln\biggl(-\frac{A_{li}\tau_{k}\tau_{ij}\tau_{li}\tau_{jkl}}
                                                    {A_{ki}\tau_{i}\tau_{kl}\tau_{jl}\tau_{ijk}}\biggr)\biggr)^2
                 -\frac{1}{2}\biggl(\ln\biggl(\frac{A_{jl}\tau_{i}\tau_{jk}\tau_{jl}\tau_{kli}}
                                                   {A_{ij}\tau_{j}\tau_{kl}\tau_{li}\tau_{ijk}}\biggr)\biggr)^2
                 +2\pi^2
\eea

Using \eqref{flip} on all the terms in the dotted boxes, all the dilogarithm terms cancel out leaving only these logarithm terms

\bea
\fl \Gamma_2 & = & +\frac{1}{2}\biggl(\ln\biggl(\frac{A_{ki}\tau_{jl}\tau_{kli}}{A_{ij}\tau_{kl}\tau_{lij}}\biggr)\biggr)^2
                 +\frac{1}{2}\biggl(\ln\biggl(-\frac{A_{jl}\tau_{k}\tau_{jl}}{A_{jk}\tau_{l}\tau_{jk}}\biggr)\biggr)^2
                 +\frac{1}{2}\biggl(\ln\biggl(-\frac{A_{ki}\tau_{jl}\tau_{ijk}}{A_{kl}\tau_{ij}\tau_{jkl}}\biggr)\biggr)^2\nn\\
\fl           && +\frac{1}{2}\biggl(\ln\biggl(\frac{A_{jl}\tau_{i}\tau_{jl}}{A_{li}\tau_{j}\tau_{li}}\biggr)\biggr)^2
                 +\frac{1}{2}\biggl(\ln\biggl(\frac{A_{ij}\tau_{kl}\tau_{lij}}{A_{jk}\tau_{li}\tau_{jkl}}\biggr)\biggr)^2
                 +\frac{1}{2}\biggl(\ln\biggl(\frac{A_{jk}\tau_{l}\tau_{jk}}{A_{kl}\tau_{j}\tau_{kl}}\biggr)\biggr)^2\nn\\
\fl           && +\frac{1}{2}\biggl(\ln\biggl(\frac{A_{kl}\tau_{ij}\tau_{jkl}}{A_{li}\tau_{jk}\tau_{lij}}\biggr)\biggr)^2
                 +\frac{1}{2}\biggl(\ln\biggl(\frac{A_{li}\tau_{j}\tau_{li}}{A_{ij}\tau_{l}\tau_{ij}}\biggr)\biggr)^2
                 +\frac{1}{2}\biggl(\ln\biggl(\frac{A_{jk}\tau_{li}\tau_{jkl}}{A_{ki}\tau_{jl}\tau_{kli}}\biggr)\biggr)^2\nn\\  
\fl           && +\frac{1}{2}\biggl(\ln\biggl(-\frac{A_{kl}\tau_{j}\tau_{kl}}{A_{jl}\tau_{k}\tau_{jl}}\biggr)\biggr)^2
                 +\frac{1}{2}\biggl(\ln\biggl(-\frac{A_{li}\tau_{jk}\tau_{lij}}{A_{ki}\tau_{jl}\tau_{ijk}}\biggr)\biggr)^2
                 +\frac{1}{2}\biggl(\ln\biggl(\frac{A_{ij}\tau_{l}\tau_{ij}}{A_{jl}\tau_{i}\tau_{jl}}\biggr)\biggr)^2\nn\\
\fl           && -\frac{1}{2}\biggl(\ln\biggl(\frac{A_{ki}\tau_{j}\tau_{li}\tau_{ki}\tau_{jkl}}
                                                   {A_{ij}\tau_{i}\tau_{jk}\tau_{kl}\tau_{lij}}\biggr)\biggr)^2
                 -\frac{1}{2}\biggl(\ln\biggl(-\frac{A_{jk}\tau_{l}\tau_{ij}\tau_{jk}\tau_{kli}}
                                                    {A_{jl}\tau_{j}\tau_{kl}\tau_{ki}\tau_{lij}}\biggr)\biggr)^2\nn\\
\fl           && -\frac{1}{2}\biggl(\ln\biggl(-\frac{A_{ki}\tau_{l}\tau_{jk}\tau_{ki}\tau_{lij}}
                                                    {A_{kl}\tau_{k}\tau_{ij}\tau_{li}\tau_{jkl}}\biggr)\biggr)^2
                 -\frac{1}{2}\biggl(\ln\biggl(\frac{A_{li}\tau_{j}\tau_{kl}\tau_{li}\tau_{ijk}}
                                                   {A_{jl}\tau_{l}\tau_{ij}\tau_{ki}\tau_{jkl}}\biggr)\biggr)^2\nn\\
\fl           && -\frac{1}{2}\biggl(\ln\biggl(\frac{A_{ij}\tau_{k}\tau_{ij}\tau_{jl}\tau_{kli}}
                                                   {A_{jk}\tau_{j}\tau_{li}\tau_{ki}\tau_{jkl}}\biggr)\biggr)^2
                 -\frac{1}{2}\biggl(\ln\biggl(\frac{A_{kl}\tau_{j}\tau_{kl}\tau_{ki}\tau_{lij}}
                                                   {A_{jk}\tau_{k}\tau_{li}\tau_{jl}\tau_{ijk}}\biggr)\biggr)^2\nn\\                                    \fl           && -\frac{1}{2}\biggl(\ln\biggl(\frac{A_{kl}\tau_{i}\tau_{kl}\tau_{jl}\tau_{ijk}}
                                                   {A_{li}\tau_{l}\tau_{jk}\tau_{ki}\tau_{lij}}\biggr)\biggr)^2
                 -\frac{1}{2}\biggl(\ln\biggl(\frac{A_{ij}\tau_{l}\tau_{ij}\tau_{ki}\tau_{jkl}}
                                                   {A_{li}\tau_{i}\tau_{jk}\tau_{jl}\tau_{kli}}\biggr)\biggr)^2\nn\\
\fl           && -\frac{1}{2}\biggl(\ln\biggl(\frac{A_{jk}\tau_{i}\tau_{jk}\tau_{kl}\tau_{lij}}
                                                   {A_{ki}\tau_{k}\tau_{ij}\tau_{jl}\tau_{kli}}\biggr)\biggr)^2
                 -\frac{1}{2}\biggl(\ln\biggl(-\frac{A_{jl}\tau_{k}\tau_{li}\tau_{jl}\tau_{ijk}}
                                                    {A_{kl}\tau_{l}\tau_{ij}\tau_{jk}\tau_{kli}}\biggr)\biggr)^2\nn\\
\fl           && -\frac{1}{2}\biggl(\ln\biggl(-\frac{A_{li}\tau_{k}\tau_{ij}\tau_{li}\tau_{jkl}}
                                                    {A_{ki}\tau_{i}\tau_{kl}\tau_{jl}\tau_{ijk}}\biggr)\biggr)^2
                 -\frac{1}{2}\biggl(\ln\biggl(\frac{A_{jl}\tau_{i}\tau_{jk}\tau_{jl}\tau_{kli}}
                                                   {A_{ij}\tau_{j}\tau_{kl}\tau_{li}\tau_{ijk}}\biggr)\biggr)^2\nn\\
\fl           && +\frac{1}{2}\biggl(\ln\biggl(-\frac{\tau_{k}\tau_{ij}\tau_{li}\tau_{jkl}}{\tau_{i}\tau_{jk}\tau_{kl}\tau_{lij}}\biggr)\biggr)^2 
                 +\frac{1}{2}\biggl(\ln\biggl(-\frac{\tau_{j}\tau_{kl}\tau_{li}\tau_{ijk}}{\tau_{l}\tau_{ij}\tau_{jk}\tau_{kli}}\biggr)\biggr)^2\nn\\
\fl           && +\frac{1}{2}\biggl(\ln\biggl(-\frac{\tau_{j}\tau_{kl}\tau_{ki}\tau_{lij}}{\tau_{k}\tau_{ij}\tau_{jl}\tau_{kli}}\biggr)\biggr)^2
                 +\frac{1}{2}\biggl(\ln\biggl(-\frac{\tau_{l}\tau_{jk}\tau_{ki}\tau_{lij}}{\tau_{k}\tau_{li}\tau_{jl}\tau_{ijk}}\biggr)\biggr)^2\nn\\
\fl           && +\frac{1}{2}\biggl(\ln\biggl(-\frac{\tau_{l}\tau_{ij}\tau_{ki}\tau_{jkl}}{\tau_{i}\tau_{kl}\tau_{jl}\tau_{ijk}}\biggr)\biggr)^2
                 +\frac{1}{2}\biggl(\ln\biggl(-\frac{\tau_{i}\tau_{jk}\tau_{jl}\tau_{kli}}{\tau_{j}\tau_{li}\tau_{ki}\tau_{jkl}}\biggr)\biggr)^2
                 +3\pi^2
\eea

This simplifies to 
\bea\label{Gamma2}
\fl \Gamma_2 & = & \frac{1}{2}\bigl(-(\ln(\tau_{ijk}))^2+(\ln(\tau_{jkl}))^2-(\ln(\tau_{kli}))^2+(\ln(\tau_{lij}))^2\nn\\
\fl      && \;\;\;\;\;                        -(\ln(\tau_{i}))^2+(\ln(\tau_{j}))^2-(\ln(\tau_{k}))^2+(\ln(\tau_{l}))^2\bigr)\nn\\
\fl      && +\ln(\tau_{ijk})\ln(\tau_{kli})-\ln(\tau_{jkl})\ln(\tau_{lij})+\ln(\tau_{i})\ln(\tau_{k})-\ln(\tau_{j})\ln(\tau_{l})\nn\\
\fl      && +\ln(\tau_{ijk})\ln\biggl(-\frac{A_{jl}A_{ki}\tau_{jl}\tau_{ki}}{A_{ij}A_{jk}\tau_{kl}\tau_{li}}\biggr)
            +\ln(\tau_{jkl})\ln\biggl(\frac{A_{jk}A_{kl}\tau_{ij}\tau_{li}}{A_{jl}A_{ki}\tau_{jl}\tau_{ki}}\biggr)\nn\\
\fl      && +\ln(\tau_{kli})\ln\biggl(-\frac{A_{jl}A_{ki}\tau_{jl}\tau_{ki}}{A_{kl}A_{li}\tau_{ij}\tau_{jk}}\biggr)
            +\ln(\tau_{lij})\ln\biggl(\frac{A_{ij}A_{li}\tau_{jk}\tau_{kl}}{A_{jl}A_{ki}\tau_{jl}\tau_{ki}}\biggr)\nn\\
\fl      && +\ln(\tau_{i})\ln\biggl(\frac{A_{jl}A_{ki}\tau_{jl}\tau_{ki}}{A_{jk}A_{kl}\tau_{jk}\tau_{kl}}\biggr)
            +\ln(\tau_{j})\ln\biggl(-\frac{A_{kl}A_{li}\tau_{kl}\tau_{li}}{A_{jl}A_{ki}\tau_{jl}\tau_{ki}}\biggr)\nn\\
\fl      && +\ln(\tau_{k})\ln\biggl(\frac{A_{jl}A_{ki}\tau_{jl}\tau_{ki}}{A_{ij}A_{li}\tau_{ij}\tau_{li}}\biggr)
            +\ln(\tau_{l})\ln\biggl(-\frac{A_{ij}A_{jk}\tau_{ij}\tau_{jk}}{A_{jl}A_{ki}\tau_{jl}\tau_{ki}}\biggr)\nn\\
\fl      && +\ln(\tau_{ij})\ln\biggl(-\frac{A_{jk}}{A_{li}}\biggr)+\ln(\tau_{jk})\ln\biggl(-\frac{A_{ij}}{A_{kl}}\biggr)
            +\ln(\tau_{kl})\ln\biggl(-\frac{A_{li}}{A_{jk}}\biggr)\nn\\
\fl      && +\ln(\tau_{li})\ln\biggl(-\frac{A_{kl}}{A_{ij}}\biggr)
            +\ln\biggl(\frac{\tau_{ki}}{\tau_{jl}}\biggr)\ln\biggl(\frac{A_{ij}A_{jk}A_{kl}A_{li}\tau_{ij}\tau_{jk}\tau_{kl}\tau_{li}}
                                                                        {A_{jl}^2A_{ki}^2\tau_{jl}^2\tau_{ki}^2}\biggr).
\eea
The reader can easily check that adding \eqref{Gamma2} to $\Gamma_1$ from \eqref{Gamma1} gives zero, verifying the closure relation.

\section*{Acknowledgments}
We are grateful to P. Hydon and V. Papageorgiou for useful discussions. SBL was supported by the UK Engineering and Physical Sciences Research Council (EPSRC). GRWQ's research is supported by the Australian Research Council through the Centre for Mathematics and Statistics of Complex Systems (MASCOS). This work was completed at the Isaac Newton Institute for Mathematical Sciences, Cambridge, during the programme Discrete Integrable Systems.

\end{document}